\documentclass[aip,apl,reprint,superscriptaddress]{revtex4-1}
\usepackage{graphicx}
\usepackage{dcolumn}
\usepackage{bm}
\usepackage{natbib}
\usepackage{nicefrac}
\usepackage{color}
\usepackage[T1]{fontenc}
\begin{document}

\title{Large linear magnetoresistance in a transition-metal stannide $\beta$-RhSn$_4$}

\author{X. Z. Xing}
\affiliation{Department of Physics, Changshu Institute of Technology, Changshu 215500, People's Republic of China}
\affiliation{Department of Physics and Key Laboratory of MEMS of the Ministry of Education, Southeast University, Nanjing 211189, China}
\author{C. Q. Xu}
\affiliation{Department of Physics, Changshu Institute of Technology, Changshu 215500, People's Republic of China}
\affiliation{Department of Physics and Hangzhou Key Laboratory of Quantum Matters, Hangzhou Normal University, Hangzhou 310036, China}
\author{N. Zhou}
\affiliation{Department of Physics and Key Laboratory of MEMS of the Ministry of Education, Southeast University, Nanjing 211189, China}
\author{B. Li}
\affiliation{Information Physics Research Center, Nanjing University of Posts and Telecommunications, Nanjing, 210023, China}
\author{Jinglei Zhang}
\affiliation{High Magnetic Field Laboratory, Chinese Academy of Sciences, Hefei 230031, China}
\author{Z. X. Shi}\email{zxshi@seu.edu.cn}
\affiliation{Department of Physics and Key Laboratory of MEMS of the Ministry of Education, Southeast University, Nanjing 211189, China}
\author{Xiaofeng Xu}\email{xiaofeng.xu@hznu.edu.cn}
\affiliation{Department of Physics, Changshu Institute of Technology, Changshu 215500, People's Republic of China}
\affiliation{Department of Physics and Hangzhou Key Laboratory of Quantum Matters, Hangzhou Normal University, Hangzhou 310036, China}

\begin{abstract}
Materials exhibiting large magnetoresistance may not only be of fundamental research interest, but also can lead to wide-ranging applications in magnetic sensors and switches. Here we demonstrate a large linear-in-field magnetoresistance, $\Delta \rho/\rho$ reaching as high as $\sim$600$\%$ at 2 K under a 9 Tesla field, in the tetragonal phase of a transiton-metal stannide $\beta$-RhSn$_4$. Detailed analyses show that its magnetic responses are overall inconsistent with the classical model based on the multiple electron scattering by mobility fluctuations in an inhomogenous conductor, but rather in line with the quantum effects due to the presence of Dirac-like dispersions in the electronic structure. Our results may help guiding the future quest for quantum magnetoresistive materials into the family of stannides, similar to the role played by PtSn$_4$ with topological node arcs.
\end{abstract}


\maketitle


The desire to maximize the sensitivity of magnetic sensor or magnetic storage devices has persistently motivated the discovery of new materials manifesting large magnetoresistance (MR). For the majority of materials, however, MR is small and usually saturates at fields comparable to the inverse of mobility (B$\sim$1/$\mu$), i.e., an order of 1 T. Most notable exceptions include the silver chalcogenides\cite{Xu97,Littlewood,Rosenbaum02}, which exhibit an unusually large, non-saturating MR over a broad range of temperatures and fields, as well as recently discovered WTe$_2$, LaSb, NbP semimetals which have extremely large MR and as such, have triggered extensive research on the origin of their remarkable magnetoresistive responses\cite{Cava,Tafti,Yan15}. Among these materials of large MR, those exhibiting linear MR are of particular interest as it may be linked to the peculiar electronic states, known as Dirac Fermions/dispersions\cite{Abrikosov-PRB,Abrikosov-epl}. For example, large linear MR seen in the semimetal Cd$_3$As$_2$ is likely due to the linear dispersion across a 3D Dirac point\cite{Ong-Cd3As2,SYLi,Coldea15}. The linear MR has thus far been observed in a series of materials involving semimetals, topological insulators and narrow band-gap semiconductors\cite{Patane-NC}, in contrast to the quadratic MR often seen in standard metals. The linear MR may originate from either a classical process or quantum effects. In the classical model, linear MR was explained in terms of a random resistor network picture, originally proposed to account for the linear MR in silver chalcogenides\cite{Littlewood}. The microscopic illustration of this picture was more recently demonstrated by the distorted electron trajectories subjected to multiple scattering by low-mobility impurities in an overall high-mobility conductor\cite{Patane-NC}. A crossover from quadratic to linear $B$-dependence of MR was predicted to occur at a characteristic field $B_c$=$\mu^{-1}$, where $\mu$ is the spatially-averaged mobility\cite{Patane-NC}. This model also predicted that the size of MR would counter-intuitively increase when the fraction of impurities that act to deflect the electron motion, increases. On the other hand, linear MR may come from the quantum effects when all carriers are degenerate into the lowest Landau level (LL), i.e., in the quantum limit regime\cite{Abrikosov-PRB}. The quantum limit may be readily achieved by a moderate field in a Dirac system as its zeroth and first LLs are well separated. As a result, MR is quadratic at low fields and becomes linear-in-$B$ once the field exceeds a critical field $B_c$ such that all states occupy the lowest LL in the quantum limit. These two different mechanisms for the linear MR are distinguishable thanks to their distinct scaling of crossover field $B_c$ and the contrasting impurity dependence of MR.

Earlier studies revealed the Dirac cones in an elemental topological insulator $\alpha$-Sn\cite{Sn-13}, on which the efficient spin to charge current conversion can be achieved by spin pumping\cite{Fert16}. More recent discovery of Dirac node arcs in the highly magnetoresistive compound PtSn$_4$ by ARPES has attracted a lot of interest and motivated parallel search for Dirac states in pertinent stannides\cite{Canfield12,Canfield16}. In this article, we demonstrate this approach is feasible and the tetragonal phase of RhSn$_4$ may represent another sample to harbor Dirac states in a binary stannide, evidenced by a large linear MR which is incompatible with the classical description for the linear MR. In PtSn$_4$, unlike 3D Dirac/Weyl semimetals or Dirac nodal-line states, Dirac nodes only extend in a short line in the momentum space, thus forming a novel node arc structure\cite{Canfield16}. Unlike PtSn$_4$ that crystallizes in an orthorhombic structure (Ccca, No. 68), $\beta$-RhSn$_4$ forms a tetragonal unit cell (I4$_1$/acd), as shown in Fig. 1\cite{thesis}. The high-quality single crystals of $\beta$-RhSn$_4$ allow us to determine both the magnitude and the form of MR that we found are overall more consistent with the quantum origin for the linear MR in this material.

Accurately weighted amounts of high purity Rh powder (4N) and Sn grains (4N) were mixed in a molar ratio of Rh:Sn=1:20 (6g in total weight), sealed in an evacuated quartz tube, and then heated up to 1050$^\circ$C. Subsequently, the furnace was slowly cooled down to 650$^\circ$C over 7 days and followed by the furnace cooling. The excess amount of Sn flux was dissolved by dilute hydrochloric acid. Large pieces of dark-gray layered single crystals with typical dimensions (5.0$\times$0.8$\times$0.8) mm$^3$ were harvested.

The actual chemical compositions of the crystals were determined by energy dispersive x-ray (EDX) spectrometry (Supplementary Information). The off-stoichiometry of the as-grown samples was verified to be very small (within 1$\%$). The single crystal x-ray diffraction (XRD) was performed at room temperature using a Rigaku diffractometer with Cu $K$$\alpha$ radiation and a graphite monochromator. Measurements of MR were conducted under various field/current configurations on over 4 samples from different batches. For \textit{all} field-current configurations (some of data are not shown in the paper), linear MR was observed. Hall effect was performed by reversing the field direction and antisymmetrizing the data.


\begin{figure}
\includegraphics[width=8cm,keepaspectratio=true]{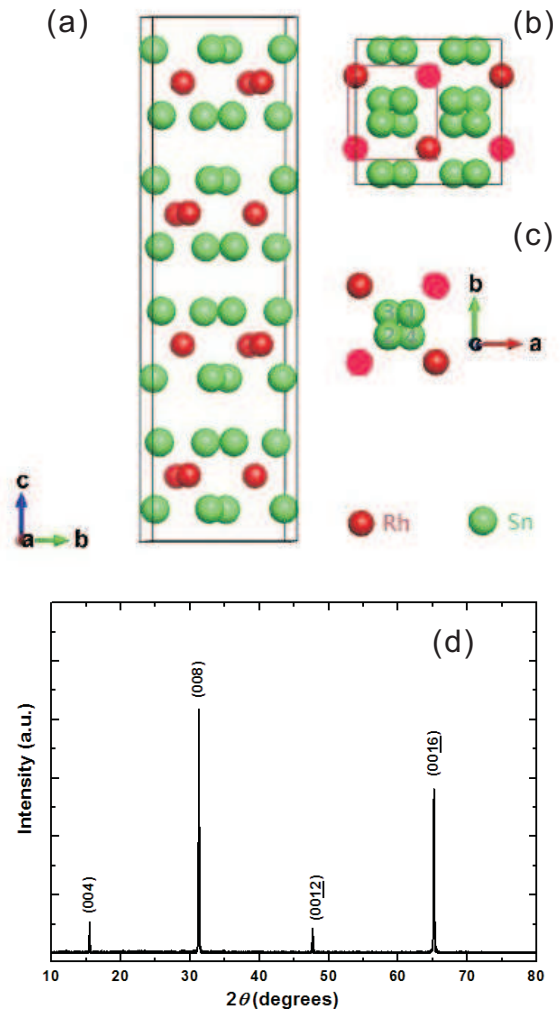}
\caption{(Color online) The crystal structure (a)-(c) and single crystal XRD (d) of $\beta$-RhSn$_4$. (b) is a perspective along the $c$-axis. (c) is the zoomed image of the red box in (b) and the number indexed on Sn atoms indicates the order in which they appear along the $c$-axis.} \label{Fig1}
\end{figure}

\begin{figure}
\includegraphics[width=8cm,keepaspectratio=true]{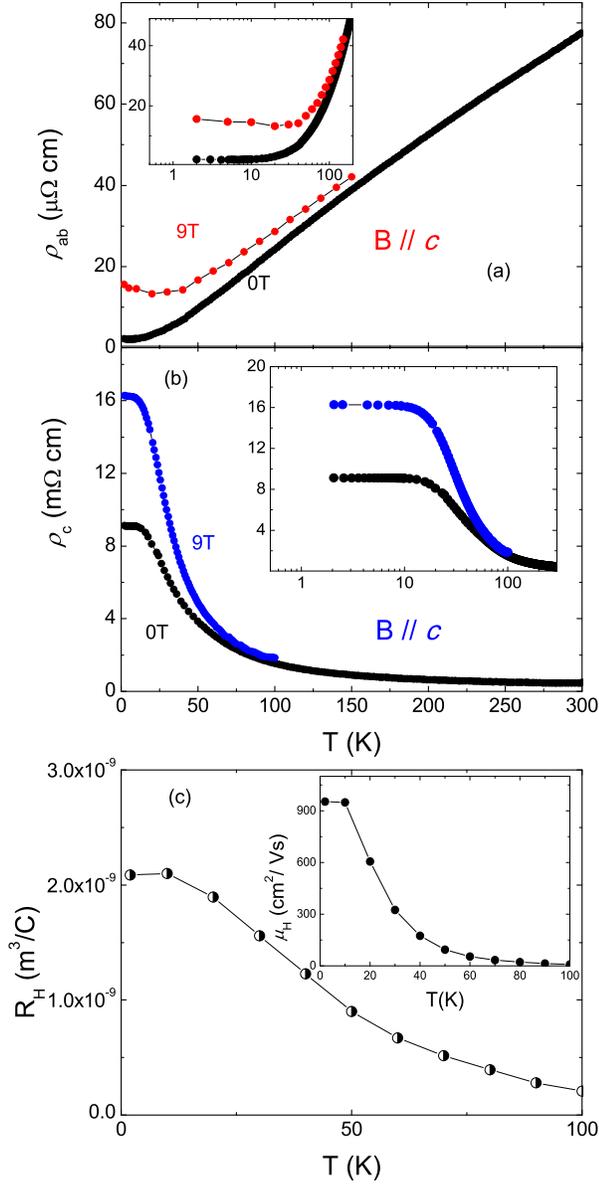}
\caption{(Color online) The color-coded in-plane (a) and inter-plane (b) resistivity under zero and 9 T field. The insets are the semi-log plots of the data. In both cases, the field is oriented along the $c$-axis. (c) The $T$-dependence of Hall coefficient below 100 K. The inset is the Hall mobility derived from $\mu_H$=$R_H$/$\rho_{ab}$.} \label{Fig2}
\end{figure}

The schematic of one unit cell in $\beta$-RhSn$_4$ is presented in Fig. 1(a). Clearly, its structure consists of stacked Rh layers alternating with Sn bilayers, resulting in a 2D layered structure. However, the coordinations of the constituent atoms are rather complex. Note first that each Rh layer is rotated by 180$^\circ$ with respect to its adjacent Rh layers, and the coordinates of Sn atoms in each layer are also different from its neighboring Sn layers. This is better visualized from a perspective projected onto the $ab$ plane (Fig. 1(b)). The deep-red atoms indicate the Rh layer above the other (pale-red). Panel (c) zooms in the image of the red box in (b). The indexing number on each Sn atom indicates the order in which it appears along the $c$-axis (pointing inside). Fig. 1(d) gives the single-crystal XRD patterns of $\beta$-RhSn$_4$. Only (00$\ell$) XRD peaks are detected, indicating the good $c$-axis orientation of the as-grown samples. The XRD patterns can be well indexed based on a tetragonal cell structure with I4$_1$/acd (No. 142) space group and the calculated $c$-axis lattice constant 22.8 {\AA} is consistent with the previous report\cite{thesis}.

The panels (a)-(c) of Fig. 2 show, in sequence: the in-plane resistivity $\rho_{ab}$; the inter-plane resistivity $\rho_c$; the $T$-evolution of the Hall coefficient $R_H(T)$ below 100 K. Upon cooling from room temperature, $\rho_{ab}$ is metallic down to the lowest temperature 2 K, whereas $\rho_c$ increases smoothly with decreasing $T$, undergoes a significant upturn below $\sim$100 K, and finally saturates to a plateau below 10 K. For comparison, the respective 9 T resistivities are also included. Note that $\rho_{ab}$ at 9 T undergoes a small upturn at low temperatures. The resistivity plateaus, in both $\rho_{ab}$ and $\rho_c$, can be clearly seen from the semi-log plots shown in the insets. This resistivity plateau is reminiscent of what was observed in LaSb and SmB$_6$, suggestive of possible surface states\cite{Tafti,SmB6-Wolgast}. It is well known that in quasi-two-dimensional materials, the hopping integral normal to the two-dimensional conducting layer $t_\perp$ is small. When the in-plane scattering are sufficiently strong to block the coherent inter-plane wave propagation, i.e., $\hbar/\tau$$>$$t_\perp$, the inter-plane transport may become decoupled, leading to a divergent $c$-axis resistivity\cite{Kartsovnik}. The possible surface conductance in topological materials, however, will saturate the insulating bulk resistance. This speculation needs further investigations in $\beta$-RhSn$_4$. From Fig. 2(c), the Hall coefficient $R_H$ also has a marked change with $T$ and levels off at low temperatures. The resultant Hall mobility, calculated from $\mu_H$=$R_H$/$\rho_{ab}$, is plotted as the inset of Fig. 2(c). Its value at 2 K is less than 1000 cm$^2$/Vs.

\begin{figure}
\includegraphics[width=9cm,keepaspectratio=true]{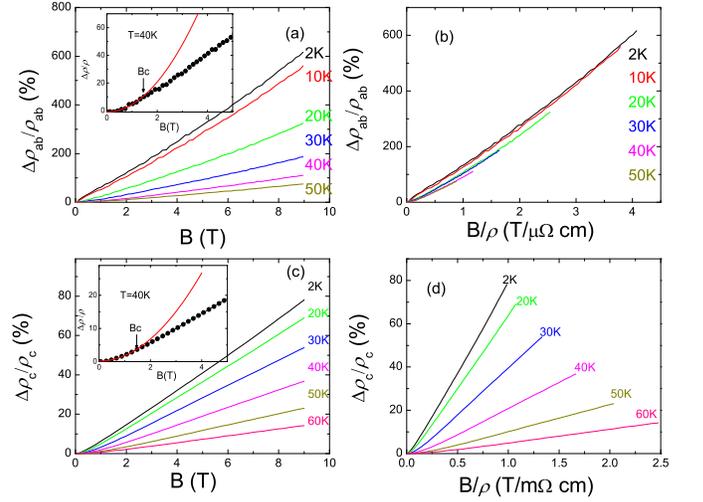}
\caption{(Color online) The in-plane MR (a) and its Kohler's plot (b). The inset of (a) zooms in the low-$B$ MR of 40 K data as an example and the red line is the $B^2$ fit below $B_c$. The corresponding plots for inter-plane transport are given in (c) and (d). The field is aligned along the $c$-axis in both cases. } \label{Fig3}
\end{figure}

\begin{figure}
\includegraphics[width=9cm,keepaspectratio=true]{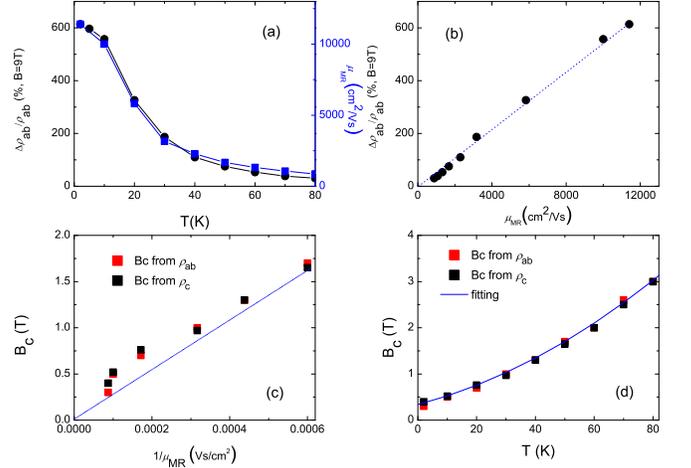}
\caption{(Color online) (a) The $T$-dependence of MR at 9 T and the effective mobility $\mu_{MR}$ (see main text). (b) shows the proportionality between MR and mobility. The blue dashed line is the linear guide to the eye. (c) The crossover field $B_c$ versus $\mu_{MR}$. Line is a guide to the eye. (d) The $T$-dependence of $B_c$. The line is the fit to $B_c$=$\frac{1}{2e\hbar v_F^2}$$(E_F+k_BT)^2$. } \label{Fig4}
\end{figure}

Fig. 3 summarizes the main magnetic responses for both the in-plane and inter-plane transport and the according Kohler's rule\cite{Kohler1938,NieLuo02}. A large linear in-plane MR, over 600$\%$ at 2 K and 9 T, is evident from Fig. 3(a). This large MR is comparable in size to the semimetals Cd$_3$As$_2$\cite{SYLi,Coldea15} and LaAgSb$_2$\cite{Petrovic12-LaAgSb} at the same temperature. However, this large MR is damped fast with increasing $T$. A close-up view of the low field behaviors, exemplified for $T$=40 K in the inset, reveals a crossover from low-$B$ quadratic dependence to high-$B$ linear MR across a critical field $B_c$ (More precisely, above $B_c$, the dominant $B$-linear MR is contaminated by a small $B^2$ component). From the low-$B$ quadratic MR, we obtain the effective mobility, denoted as $\mu_{MR}$, through $\Delta \rho/\rho$=($\mu_{MR}B$)$^2$. The as-estimated $\mu_{MR}$ exceeds 10$^5$ cm$^2$/Vs at 2 K, two orders of magnitude larger than the Hall mobility. This discrepancy in the mobility values drawn using two distinct ways is possibly because the Hall mobility $\mu_H$=$R_H$/$\rho_{ab}$ is derived from a single-band model. In two-band materials, however, the Hall coefficient $R_H$ may be significantly reduced compared with their individual coefficients for electrons ($R_e$) and holes ($R_h$), because the Hall coefficient at low fields is given by $R_H$=$\frac{\sigma_h^2 R_h +\sigma_e^2 R_e}{(\sigma_h +\sigma_e)^2}$, where $\sigma_{e(h)}$ and $R_{e(h)}$ are the conductivity and Hall coefficient of $e(h)$ carriers, respectively. Note that $R_e$ and $R_h$ are opposite in sign. The resultant Hall effect may become small due to the partial cancellation of hole and electron components. As a consequence, the Hall mobility can be substantially underestimated. In what follows, we use $\mu_{MR}$ for the analysis since it is difficult to account for the large observed MR using such a small Hall mobility. However, analysis on the footing using $\mu_H$ would not change the main results, hence the conclusion of this article (see Supplementary Information). Fig. 3(c) displays the similar behaviors observed in the inter-plane MR. As seen from Fig. 3(b) and (d), the Kohler's rule is roughly obeyed for the in-plane MR but is grossly violated for the out-of-plane transport in the regime where $\rho_c$ is insulating. To date, the Kohler's rule was found to be violated in a number of metals, including cuprates\cite{Harris95}, iron pnictides\cite{Ghosh}, as well as organic conductors\cite{McKenzie}.

Fig. 4(a) compares the $T$-dependence between the size of MR and the effective mobility. Obviously, these two quantities follow the same trend as $T$ decreases. Their linear correlation is clearly presented in Fig. 4(b). We now turn next to the origin of the linear MR observed in this material. Note that both the classical and quantum models of linear MR give $\Delta \rho/\rho$$\propto$$\mu$, exactly as seen in Fig. 4(b). In the classical model, i.e., the mobility-fluctuation-induced linear MR in an inhomogenous conductor, theory predicted the crossover field $B_c$ to be inversely proportional to the mobility, $B_c$$\sim$$\mu$$^{-1}$. In $\beta$-RhSn$_4$, as shown in Fig. 4(c), $B_c$ tends to deviate from this law, especially in the low-$T$ regime. Additionally, as the model claimed, since the low-mobility island, regarded as a scattering center, acts to reflect the current path and cause an increase of resistivity, the MR shall go up when more impurities are included in the system. However, as shown in the Supplementary Information, our measurements on a set of samples of different purity suggest otherwise. In a second quantum model, the critical field $B_c$ occurs at $B_c$=$\frac{1}{2e\hbar v_F^2}$$(E_F+k_BT)^2$ when all carriers are degenerate to the zeroth LL\cite{Petrovic11-SrMnBi,Petrovic12-LaAgSb}. We examined this scaling in our compound and found it fits to the experimental data well, yielding $E_F$=3.5 meV, $v_F$=1.64$\times$10$^5$ m/s. As a comparison, the same analysis in SrMnBi$_2$ gives $E_F$=4.9 meV and $v_F$=5.1$\times$10$^5$ m/s.\cite{Petrovic11-SrMnBi}


In summary, high-quality single crystals of $\beta$-RhSn$_4$ exhibiting large linear magnetoresistance were successfully synthesized and characterized. This large linear MR is similar in magnitude to what was seen in semimetals Cd$_3$As$_2$\cite{SYLi,Coldea15} and LaAgSb$_2$\cite{Petrovic12-LaAgSb}, and larger than in Dirac material SrMnBi$_2$\cite{Petrovic11-SrMnBi}. The mechanism for this linear MR in this stannide is more likely of quantum origin where Dirac fermions may emerge near the Fermi level. Future ARPES and band structure calculations may help to clarify the underlying physics.


See Supplementary Information for the elemental analysis, resistivity measurement, impurity dependence of MR, and scaling using Hall mobility details.

The authors would like to thank C. M. J. Andrew, A. F. Bangura, W. H. Jiao for stimulating discussions. This work is sponsored by the National Key Basic Research Program of China (Grant No. 2014CB648400), and by National Natural Science Foundation of China (Grant No. 11474080, U1432135, 11611140101). X.X. would also like to acknowledge the financial support from the Distinguished Young Scientist Funds of Zhejiang Province (LR14A040001) and an open program from Wuhan National High Magnetic Field Center (2015KF15).

\end{document}